\documentclass[12pt,a4paper]{article}
\usepackage{epsfig}
\begin{document}
\title{New gauge bosons from the littlest Higgs model \\
and the process $e^{+}e^{-}\rightarrow t\bar{t}$}
\author{Chong-Xing Yue, Lei Wang, Jian-Xing Chen\\
{\small Department of Physics, Liaoning  Normal University, Dalian
116029. P.R.China}
\thanks{E-mail:cxyue@lnnu.edu.cn}}
\date{\today}
\maketitle
\begin{abstract}
\hspace{5mm} In the context of the littlest Higgs$(LH)$ model, we
study the process $ e^{+}e^{-}\rightarrow t\bar{t}$. We find that
the new gauge bosons $Z_{H}$ and $B_{H}$ can produce significant
correction effects on this process, which can be further enhanced
by the suitably polarized beams. In most of the parameter space
preferred by the electroweak precision data, the absolute value of
the relative correction parameter $R_{B_{H}}$  is larger than
$5\%$. As long as $1TeV\leq M_{Z_{H}}\leq 1.5TeV$ and $0.3\leq
c\leq 0.5,$ the absolute value of the relative correction
parameter $R_{Z_{H}}$ is larger than $5\%$. With reasonable values
of the parameters of the $LH$ model, the possible signals of the
new gauge bosons $B_{H}$ and $Z_{H}$ can be detected via the
process $ e^{+}e^{-} \rightarrow t\bar{t}$ in the future $LC$
experiments with the c.m. energy $\sqrt{S}=800GeV$. $B_{H}$
exchange and $Z_{H}$ exchange can generate significantly
corrections to the forward-backward asymmetry $A_{FB}(t\bar{t})$
only in small part of the parameter space.
\end {abstract}
\newpage
\noindent{\bf I. Introduction}

Although the standard model($SM$) that bases on the gauge group
$SU(2)_{L}\times U(1)_{Y}$ has been successful in describing the
physics of electroweak interactions, the mechanism of the
electroweak symmetry breaking($EWSB$) and the origins of the
masses of the elementary fermions are still unknown. Furthermore,
its scalar sector suffers from the problems of triviality and
unnaturalness, etc. Thus, it is quite possible that the $SM$ is
only an effective theory valid below some high energy scale. New
physics($NP$) should exist at energy scales around $TeV$.

Recently, a kind of theory for $EWSB$ was proposed to solve the
hierarchy between the $TeV$ scale of possible $NP$ and the
electroweak scale $v=246GeV$, which is known as "little Higgs
models"[1,2,3]. The key feature of these models is that the Higgs
boson is a pseudo-Goldstone boson of a global symmetry which is
spontaneously broken at some higher scale $f$ and thus is
naturally light. $EWSB$ is induced by a Coleman-Weinberg
potential, which is generated by integrating out the heavy degrees
of freedom. This type of models can be regarded as one of the
important candidates of the $NP$ beyond the $SM$.

A high energy $e^{+}e^{-}$ linear collider($LC$) will offer an
opportunity to make precision measurement of the properties of the
electroweak gauge bosons, top quarks, Higgs bosons and also to
constrain $NP$ [4]. In the $LC$ experiments, top quark pairs are
mainly produced from the S-channel exchange of the $SM$ gauge
bosons $\gamma$ and $Z$ via the process $e^{+}e^{-}\rightarrow
t\bar{t}$ [5]. The total cross section is of the order of 1$pb$,
so that top quark pairs will be produced at large rates in a clean
environment at $LC$. If we assume that the integrated luminosity
$\pounds_{int}$ is about 100$fb^{-1}$, there will be several times
$10^{4}$ top quark pairs to be generated in the future $LC$
experiments. Furthermore, the $QCD$ and $EW$ corrections to the
process $e^{+}e^{-}\rightarrow t\bar{t}$ are small and decrease as
the centre-of-mass(c.m.) energy $\sqrt{S}$ increasing. The option
of longitudinally polarized beams can help to improve the
measurement precision and reduce background in search for $NP$.
Thus, theoretical calculations of new particles contributions to
the process $e^{+}e^{-}\rightarrow t\bar{t}$ are of much interest
for testing of $NP$.

In general, the new gauge bosons are heavier than the current
experimental limits on direct searches. However, these new
particles may produce virtual effects on some physical observable,
which may be detected in the present or future high energy
experiments. In Ref.[6], we discussed the possible of detecting
the new gauge bosons $Z_{H}$ and $B_{H}$ predicted by the littlest
Higgs(LH) model [1] in the future $LC$ experiments with the c.m.
energy $\sqrt{S}=500GeV$ and the integrating luminosity
$\pounds_{int}=340fb^{-1}$ and both beams polarized via
considering their contributions to the processes
$e^{+}e^{-}\rightarrow f\bar{f}$ with $f=\tau,\mu,b$ and $c$.
Since the masses of these fermions are largely smaller than the
c.m. energy $\sqrt{S}$, we have neglected the masses of these
fermions in our numerical estimations. Our results show that the
new gauge bosons $Z_{H}$ and $B_{H}$ can indeed produce
significant contributions to these process in most of the
parameter space preferred by the electroweak precision data, which
might be observable in the future $LC$ experiments. The aim of
this paper is to consider the contributions of the $Z_{H}$ and
$B_{H}$ to the process $e^{+}e^{-}\rightarrow t\bar{t}$ and
discuss whether these new particles can be detected via this
process in the future $LC$ experiments with the c.m. energy
$\sqrt{S}=800GeV$ and the integrating luminosity
$\pounds_{int}=580fb^{-1}$. We find that the absolute value of the
relative correction parameter $R_{B_{H}}$ generated by $B_{H}$
exchange is larger than $8\%$ in most of the parameter space of
the $LH$ model preferred by the electroweak precision data. As
long as $1TeV\leq M_{Z_{H}}\leq 1.5TeV$ and $0.3\leq c\leq 0.5,$
the absolute value of $R_{Z_{H}}$ is larger than 5\%. If we assume
that the initial electron and positron beams are suitably
polarized, the absolute values of the relative correction
parameters $R_{B_{H}}$ and $R_{Z_{H}}$ can be enhanced. Thus, with
reasonable values of the parameters of the $LH$ model, the
possible signals of the new gauge bosons $B_{H}$ and $Z_{H}$ can
be detected in the future $LC$ experiments with the c.m. energy
$\sqrt{S}=800GeV$, which is similar to the conclusions given in
Ref.[6]. We further calculate the contributions of these new gauge
bosons to the forward-backward asymmetry $A_{FB}(t\bar{t})$. We
find that they can generate significantly corrections to the
forward-backward asymmetry $A_{FB}(t\bar{t})$ only in small part
of the parameter space.

In section II, we give the formula of the contributions of new
gauge bosons $B_{H}$ and $Z_{H}$ to the process
$e^{+}e^{-}\rightarrow t\bar{t}$ and estimate the values of the
relative corrections parameters $
R_{B_{H}}=\sigma^{B_{H}}(t\bar{t})/\sigma^{SM}(t\bar{t})$ and
$R_{Z_{H}}=\sigma^{Z_{H}}(t\bar{t})/\sigma^{SM}(t\bar{t})$. The
dependence of the relative correction parameters $R_{B_{H}}$ and
$R_{Z_{H}}$ on the initial beam polarization is discussed in
section III. In section IV, we calculate the contributions of
these new gauge bosons to the forward-backward asymmetry
$A_{FB}$$(t\bar{t}).$ Our conclusions and discussions are given in
section V.

\noindent{\bf II. Corrections of the new gauge bosons $B_{H}$ and
$Z_{H}$ to the process $e^{+}e^{-}\rightarrow t\bar{t}$ }

The $LH$ model [1] is one of the simplest and phenomenologically
viable models, which realizes the little Higgs idea. It consists
of a non-linear $\sigma$ model with a global $SU(5)$ symmetry,
which is broken down to its subgroup $SO(5)$ by a vacuum
condensate $f\sim\Lambda s/4\pi\sim TeV$. At the same time, the
locally gauged group $SU(2)_{1}\times U(1)_{1}\times
SU(2)_{2}\times U(1)_{2}$ is broken to its diagonal subgroup
$SU(2)\times U(1)$, identified as the $SM$ electroweak gauge
group. This breaking scenario gives rise to four massive gauge
bosons $B_{H}$, $Z_{H}$, and $W_{H}^{\pm}$, which might produce
characteristic signatures at the present and future high energy
collider experiments [7,8,9].

Taking account of the gauge invariance of the Yukawa coupling and
the $U(1)$ anomaly cancellation, the coupling expressions of the
gauge bosons $B_{H}$ and $Z_{H}$ to ordinary particles, which are
related to our calculation, can be written as [7]:
\begin{eqnarray}
g_{V}^{B_{H}ee}&=&\frac{3e}{4C_{w}s'c'}(c'^{2}-\frac{2}{5}),\hspace{2.0cm}
g_{A}^{B_{H}ee}=\frac{e}{4C_{w}s'c'}(c'^{2}-\frac{2}{5});\\
g_{V}^{B_{H}tt}&=&\frac{e}{2C_{w}s'c'}[\frac{5}{6}(\frac{2}{5}-c'^{2})-\frac{1}{5}x_{L}],
\hspace{0.15cm}g_{A}^{B_{H}tt}=\frac{e}{2C_{w}s'c'}[\frac{1}{2}(\frac{2}{5}-c'^{2})-
\frac{1}{5}x_{L}];\\g_{V}^{Z_{H}ee}&=&-\frac{ec}{4S_{w}s},\hspace{3.6cm}
g_{A}^{Z_{H}ee}=\frac{ec}{4S_{w}s};\\
g_{V}^{Z_{H}tt}&=&\frac{ec}{4S_{w}s},\hspace{3.9cm}
g_{A}^{Z_{H}tt}=-\frac{ec}{4S_{w}s}.
\end{eqnarray}
Where $S_{w}=sin\theta_{w}, \theta_{w}$ is the Weinberg angle.
Using the mixing parameters $c(s=\sqrt{1-c^{2}})$ and
$c'(s'=\sqrt{1-c'^{2}})$, we can represent the $SM$ gauge coupling
constants as $g=g_{1}s=g_{2}c$ and $g'=g'_{1}s'=g'_{2}c'$. The
mixing angle parameter between the $SM$ top quark t and the
vector-like quark T is defined as
$x_{L}=\lambda_{1}^{2}/(\lambda_{1}^{2}+\lambda_{2}^{2})$, in
which $\lambda_{1}$ and $\lambda_{2}$ are the Yukawa coupling
parameters.

Global fits to the eletroweak precision data produce rather severe
constraints on the parameter space of the $LH$ model [10].
However, if the $SM$ fermions are charged under $U(1)_{1}\times
U(1)_{2},$ the constraints become relaxed. The scale parameter
$f=1\sim 2TeV$ is allowed for the mixing parameters $c$, $c',$ and
$x_{L}$ in the ranges of $0\sim 0.5$, $0.62\sim 0.73$, and
$0.3\sim0.6$, respectively [11]. In this case, the masses of
$B_{H}$ and $Z_{H}$ are allowed in the ranges of $300GeV\sim
900GeV$ and $1TeV\sim 3TeV$, respectively. Thus, we will take the
$Z_{H}$ mass $M_{Z_{H}}$, $B_{H}$ mass $M_{B_{H}}$ and the mixing
parameters $c,$ $c'$ and $x_{L}$ as free parameters in our
calculation.

\begin{figure}[htb]
\vspace{0cm}
  \centering
   \includegraphics[width=3.3in]{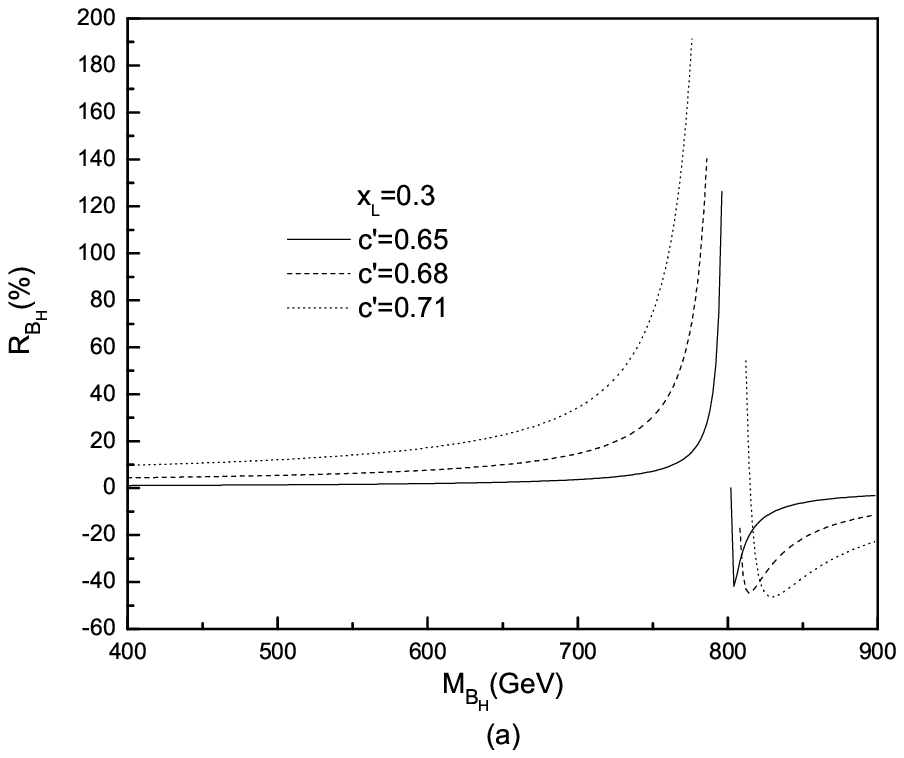}
    \hspace{-0.4in}
   \includegraphics[width=3.3in]{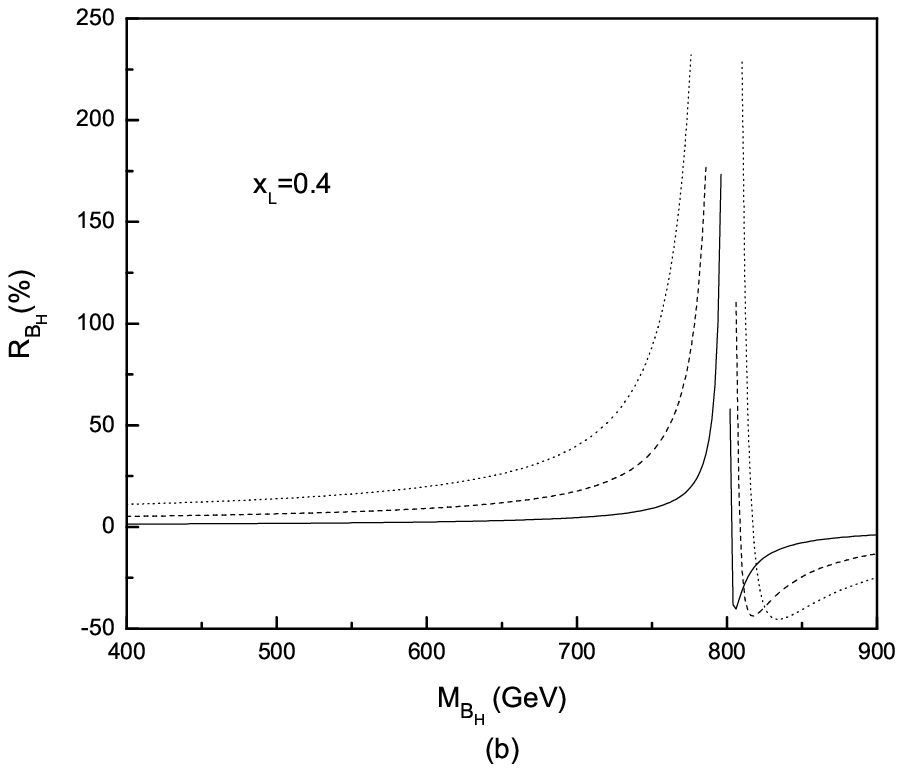}
\end{figure}

\begin{figure}[htb]
\vspace{-1.5cm}
  \centering
   \includegraphics[width=3.3in]{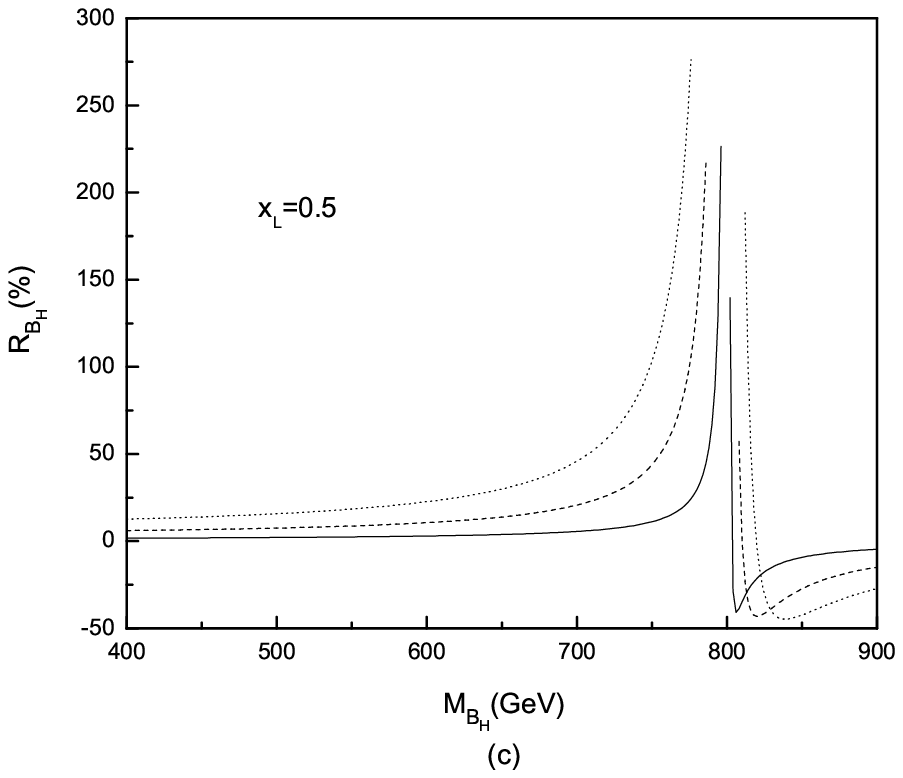}
    \hspace{-0.4in}
   \includegraphics[width=3.3in]{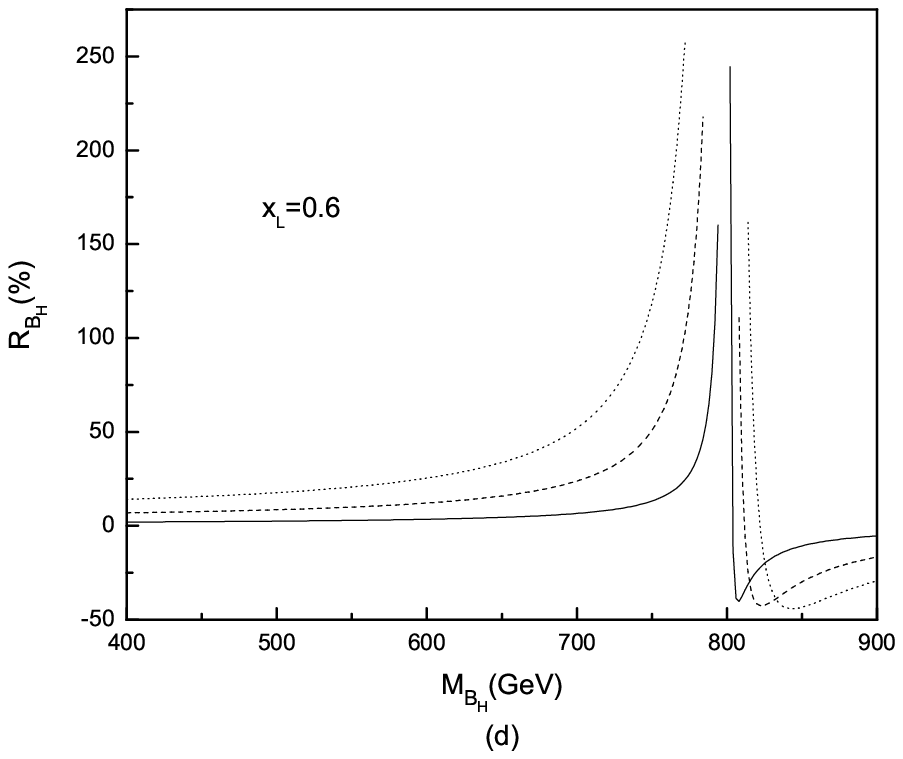}
  \caption{The relative correction parameter
$R_{B_{H}}$ as a function of the $B_{H}$ mass $M_{B_{H}}$ for
\hspace*{1.8cm} different values of the mixing parameters $c'$ and
$x_{L}$. }
\end{figure}

For the $SM$, top quark pair $t\bar{t}$ can be produced in
sufficient abundance in the $LC$ experiments. The main production
mechanism proceed at the Born level by the S-channel annihilation
of an initial electron-position pair into virtual photon or
neutral gauge boson $Z$, and their subsequent splitting into top
quark pairs, $e^{+}e^{-}\rightarrow\gamma,Z\rightarrow t\bar{t}.$
For the $LH$ model, the $B_{H}$ exchange and $Z_{H}$ exchange can
also produce the top quark pairs. The production cross sections
can be written as:
\begin{eqnarray}
\sigma^{B_{H}}(t\bar{t})&=&\frac{N_{c}^{f}\beta}{8\pi S}
\{(1-\frac{\beta^{2}}{3})\frac{4}{3}e^{2}g_{V}^{B_{H}ee}g_{V}^{B_{H}tt}\frac{S
(M_{B_{H}}^{2}-S)}
{(S-M_{B_{H}}^{2})^{2}+M_{B_{H}}^{2}\Gamma_{B_{H}}^{2}}\nonumber\\
&&+[(g_{V}^{B_{H}ee})^{2}+(g_{A}^{B_{H}ee})^{2}][(1-\frac{\beta^{2}}{3})
[(g_{V}^{B_{H}tt})^{2}
+(g_{A}^{B_{H}tt})^{2}]-(1-\beta^{2})(g_{A}^{B_{H}tt})^{2}]\nonumber\\&&\frac{S^{2}}
{(S-M_{B_{H}}^{2})^{2}+M_{B_{H}}^{2}\Gamma_{B_{H}}^{2}}+(g_{V}^{Zee}
g_{V}^{B_{H}ee}+g_{A}^{Zee}g_{A}^{B_{H}ee})\nonumber\\&&[(1-\frac{\beta^{2}}{3})
(g_{V}^{Ztt}g_{V}^{B_{H}tt}+g_{A}^{Ztt}g_{A}^{B_{H}tt})-(1-\beta^{2})(g_{A}^{B_{H}tt})
(g_{A}^{Ztt})]\nonumber\\&&\frac{2S^{2}[(S-M_{Z}^{2})(S-M_{B_{H}}^{2})+M_{Z}
\Gamma_{Z}M_{B_{H}}\Gamma_{B_{H}}]}{[(S-M_{Z}^{2})^{2}+M_{Z}^{2}\Gamma_{Z}^{2}]
[(S-M_{B_{H}}^{2})^{2}+M_{B_{H}}^{2}\Gamma_{B_{H}}^{2}]}\},\\
\sigma^{Z_{H}}(t\bar{t})&=&\frac{N_{c}^{f}\beta}{8\pi S}
\{(1-\frac{\beta^{2}}{3})\frac{4}{3}e^{2}g_{V}^{Z_{H}ee}g_{V}^{Z_{H}tt}\frac{S
(M_{Z_{H}}^{2}-S)}
{(S-M_{Z_{H}}^{2})^{2}+M_{Z_{H}}^{2}\Gamma_{Z_{H}}^{2}}\nonumber\\
&&+[(g_{V}^{Z_{H}ee})^{2}+(g_{A}^{Z_{H}ee})^{2}][(1-\frac{\beta^{2}}{3})
[(g_{V}^{Z_{H}tt})^{2}
+(g_{A}^{Z_{H}tt})^{2}]-(1-\beta^{2})(g_{A}^{Z_{H}tt})^{2}]\nonumber\\&&\frac{S^{2}}
{(S-M_{Z_{H}}^{2})^{2}+M_{Z_{H}}^{2}\Gamma_{Z_{H}}^{2}}+(g_{V}^{Zee}
g_{V}^{Z_{H}ee}+g_{A}^{Zee}g_{A}^{Z_{H}ee})\nonumber\\&&[(1-\frac{\beta^{2}}{3})
(g_{V}^{Ztt}g_{V}^{Z_{H}tt}+g_{A}^{Ztt}g_{A}^{Z_{H}tt})-(1-\beta^{2})(g_{A}^{Z_{H}tt})
(g_{A}^{Ztt})]\nonumber\\&&\frac{2S^{2}[(S-M_{Z}^{2})(S-M_{Z_{H}}^{2})+M_{Z}
\Gamma_{Z}M_{Z_{H}}\Gamma_{Z_{H}}]}{[(S-M_{Z}^{2})^{2}+M_{Z}^{2}\Gamma_{Z}^{2}]
[(S-M_{Z_{H}}^{2})^{2}+M_{Z_{H}}^{2}\Gamma_{Z_{H}}^{2}]}\}
\end{eqnarray}
with
\begin{equation}
g_{V}^{Zee}=\frac{e}{4S_{w}C_{w}}(-1+4S_{w}^{2}),\hspace{2cm}
g_{A}^{Zee}=\frac{e}{4S_{w}C_{w}}
\end{equation}
\begin{equation}
g_{V}^{Ztt}=\frac{e}{4S_{w}C_{w}}(1-\frac{8}{3}S_{w}^{2}),\hspace{2.3cm}
g_{A}^{Ztt}=\frac{e}{4S_{w}C_{w}},
\end{equation}
where $\beta=\sqrt{1-\frac{4m_{t}^{2}}{S}}$, $m_{t}$ is the top
quark mass. $\Gamma_{i}$ represent the total decay widths of the
gauge bosons $Z,Z_{H},$ and $B_{H}$. $\Gamma_{Z_{H}}$ and
$\Gamma_{B_{H}}$ have been given in Ref.[6]. From above equations,
we can see that $\sigma^{B_{H}}(t\bar{t})$ mainly dependents the
free parameters $M_{B_{H}}$, $c'$ and $x_{L}$, while
$\sigma^{Z_{H}}(t\bar{t})$ only dependents the free parameters $c$
and $M_{Z_{H}}$, which is differently from those for the process
$e^{+}e^{-}\rightarrow f\bar{f}$ with $f=\tau,\mu,b$ and $c$. In
that case, the contributions of the gauge bosons $B_{H}$ is
independent of the mixing parameter $x_{L}$. Thus, in this paper,
we will take the mixing parameters $c,c'$ and $x_{L}$ as free
parameters. Certainly, due to the mixing between the gauge bosons
$Z$ and $Z_{H}$, the $SM$ tree-level couplings $Ze\bar{e}$ and
$Zt\bar{t}$ receive corrections at the order of $v^{2}/f^{2}$,
which can also produce contributions to the production cross
section of the process $e^{+}e^{-}\rightarrow t\bar{t}$. However,
the contributions are suppressed by the factor
 $v^{4}/f^{4}$, which are smaller than those of $B_{H}$ or
$Z_{H}$. Thus, we have neglected this kind of corrections in above
equations.

\vspace{0.5cm}
\begin{figure}[htb]
\vspace{-1cm}
\begin{center}
\epsfig{file=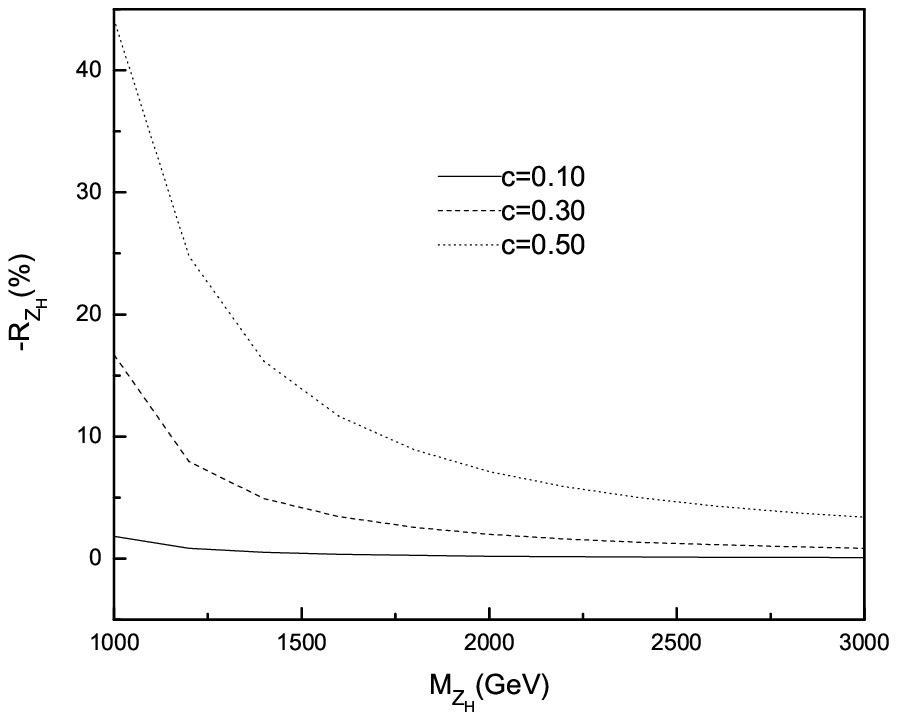,width=350pt,height=300pt} \vspace{-1cm}
\hspace{1cm} \caption{The relative correction parameter
$R_{Z_{H}}$ as a function of the $Z_{H}$ mass $M_{Z_{H}}$ for
\hspace*{1.8cm}three values of the mixing parameter $c$.}
\label{ee}
\end{center}
\end{figure}
\vspace*{0.5cm}

To see the correction effects of $B_{H}$ exchange and $Z_{H}$
exchange on the $t\bar{t}$ production cross section, we plot the
relative correction parameters
$R_{B_{H}}=\sigma^{B_{H}}(t\bar{t})/\sigma^{SM}(t\bar{t})$ and
$R_{Z_{H}}=\sigma^{Z_{H}}(t\bar{t})/\sigma^{SM}(t\bar{t})$ as
functions of $M_{B_{H}}$and$M_{Z_{H}}$ in Fig.1 and Fig.2,
respectively. From these figures, we can see that the gauge boson
$Z_{H}$ decreases the $SM$ $t\bar{t}$ production cross section
$\sigma^{SM}(t\bar{t})$ in all of the parameter space, which
satisfies the electroweak precision constraints. In most part of
the parameter space, the absolute value of the relative correction
parameter $R_{Z_{H}}$ is smaller than $5\%$, which is very
difficult to be detected in the future $LC$ experiments. This is
consistent with the contributions of $Z_{H}$ to the process
$e^{+}e^{-}\rightarrow f\bar{f}$, which has been studied in
Ref.[6]. However, for the gauge boson $B_{H}$, it is not this
case. For $M_{B_{H}}\leq800GeV$, $B_{H}$ exchange produce positive
corrections to the $t\bar{t}$ production cross section
$\sigma^{SM}(t\bar{t})$ and the value of $R_{B_{H}}$ increase as
$M_{B_{H}}$, $x_{L}$ and $c'$ increasing. For
$800GeV<M_{B_{H}}\leq900GeV$, $B_{H}$ exchange decrease the cross
section $\sigma^{SM}(t\bar{t})$ and the absolute of $R_{B_{H}}$
increase as $M_{B_{H}}$ decreasing and $x_{L}$, $c'$ increasing.
The peak of the $R_{B_{H}}$ resonance emerges when the $B_{H}$
mass $M_{B_{H}}$ is approximately  equal to the c.m. energy
$\sqrt{S}=800GeV$. In most part of the parameter space, the
absolute value of $R_{B_{H}}$ is larger than 8\%. Thus, the
virtual effects of $B_{H}$ on the process $e^{+}e^{-}\rightarrow
t\bar{t}$ should be easy detected in the future $LC$ experiment
with $\sqrt{S}=800GeV$ and $\pounds_{int}=580fb^{-1}$.

\noindent{\bf III. The dependence of the relative correction
       parameters $R_{B_{H}}$ and $R_{Z_{H}}$ on the

       electron and positron beam polarization}

An $LC$ has a large potential of the discovery of new particles
and is well suited for the precise analysis of $NP$ beyond the
$SM$. At present, the existing proposals are designed with high
luminosity of about $\pounds_{int}=340fb^{-1}$ at
$\sqrt{S}=500GeV$ and $\pounds_{int}=580fb^{-1}$ at
$\sqrt{S}=800GeV$ [4]. An important tool of an $LC$ is the use of
polarized beams. Beam polarization is not only useful for a
possible reduction of the background, but might also serve as a
possible tool to disentangle different contributions to the signal
and lead to substantial enhancement of the produce cross sections
of some processes [12]. To see whether the contributions of the
new gauge bosons $B_{H}$ and $Z_{H}$ to the process
$e^{+}e^{-}\rightarrow t\bar{t}$ can indeed be detected, we
discuss the dependence of the relative correction parameters
$R_{B_{H}}$ and $R_{Z_{H}}$ on the initial electron and positron
beam polarization in this section.

Considering the polarization of the initial electron and positron
beams, the cross section of the process $e^{+}e^{-}\rightarrow
t\bar{t}$ can be generally written as:
\begin{equation}
\sigma(t\bar{t})=(1+P_{e})(1-P_{\bar{e}})(\sigma_{RR}(t\bar{t})+\sigma_{RL}(t\bar{t}))
+(1-P_{e})(1+P_{\bar{e}})(\sigma_{LL}(t\bar{t})+\sigma_{LR}(t\bar{t})),
\end{equation}
where $P_{e}$ and $P_{\bar{e}}$ are the degrees of longitudinal
electron and position polarization, respectively. $\sigma_{ij}$
are the chiral cross sections of this process. The relative
correction parameters $R_{B_{H}}$ and $R_{Z_{H}}$ are plotted as
functions of $M_{B_{H}}$ and $M_{Z_{H}}$ for $c'=0.65$,
$x_{L}=0.5$, $c=0.3$ and different beam polarizations in Fig.3 and
Fig.4, respectively. In these two figures, we have used the solid
line, dashed line, and dotted line to represent
$(P_{e},P_{\bar{e}})$=$(0,0)$, $(0.8,-0.6)$, and $(-0.8,0.6)$,
respectively. Our calculation results show that the absolute
values of $R_{B_{H}}$[$R_{Z_{H}}$] for
$(P_{e},P_{\bar{e}})=(0.8,0.6)$[(-0.8,-0.6)] are smaller than
those for $(P_{e},P_{\bar{e}})=(0,0)$. Thus, in Fig.3 and Fig.4 we
do not plot these lines.

\vspace{0.5cm}
\begin{figure}[htb]
\vspace{-1cm}
\begin{center}
\epsfig{file=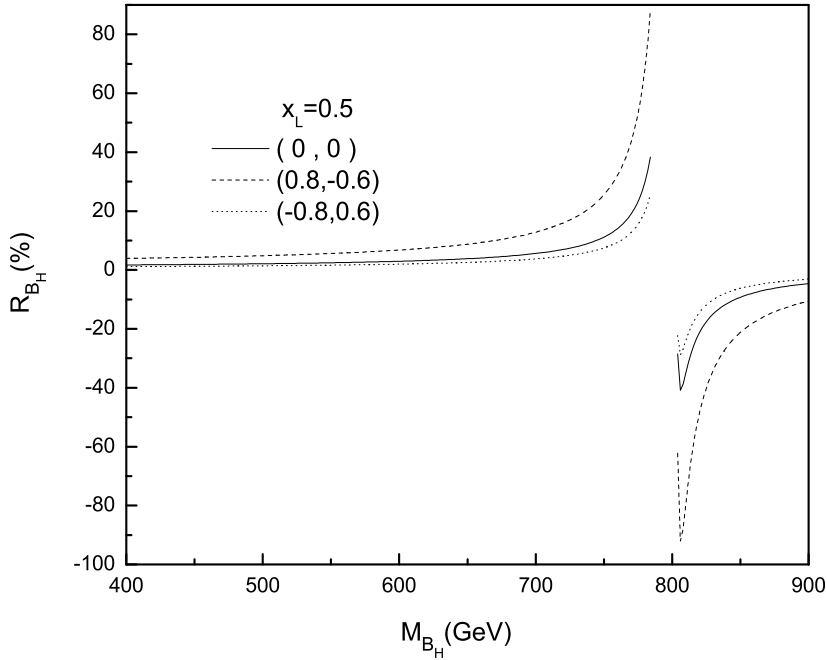,width=350pt,height=300pt} \vspace{-1cm}
\hspace{1cm} \caption{The relative correction parameter
$R_{B_{H}}$ as a function of the $B_{H}$ mass $M_{B_{H}}$ for
\hspace*{1.8cm} $c'=0.65$, $x_{L}=0.5$, and
$(P_{e},P_{\bar{e}})=(0,0), (0.8,-0.6), (-0.8,0.6)$. } \label{ee}
\end{center}
\end{figure}

From Fig.3 and Fig.4 we can see that the suitably polarized beams
can indeed enhance the virtual effects of the new gauge bosons
$B_{H}$ and $Z_{H}$ on the process $e^{+}e^{-}\rightarrow
t\bar{t}$. In the whole parameter space preferred by the
electroweak precision data, the value of $R_{B_{H}}$ for
$(P_{e},P_{\bar{e}})=(0.8,-0.6)$ is larger than that for
$(P_{e},P_{\bar{e}})=(0,0)$, while the absolute values of
$R_{Z_{H}}$ for $(P_{e},P_{\bar{e}})=(-0.8,0.6)$ is larger than
that for $(P_{e},P_{\bar{e}})=(0,0)$. Varying the values of the
free parameters $c'$, $x_{L}$, and $c$ does not change this
conclusion. So, in Fig.3 and Fig.4 we have taken these parameters
for fixed values $x_{L}=0.5,$ $c'=0.65$, and $c=0.3$. Certainly,
the values of $R_{B_{H}}$ and $R_{Z_{H}}$ change as the values of
these parameters varying. For example, for $0.3\leq c \leq 0.5$
and $1TeV\leq M_{Z_{H}}\leq 2TeV$, the absolute value of
$R_{Z_{H}}$ for $(P_{e},P_{\bar{e}})=(-0.8,0.6)$ is larger than
6\%. The absolute of $R_{B_{H}}$ for
$(P_{e},P_{\bar{e}})=(0.8,-0.6)$ is larger than $5\%$ for
$x_{L}=0.5$, $0.68\leq c'\leq 0.73$ and $500GeV<M_{B_{H}}\leq
900GeV$, but for $x_{L}=0.6$ its value is larger than 5\% for
$0.65\leq c'\leq 0.73$ and $450GeV\leq M_{B_{H}}\leq 900GeV$.
Thus, using the suitably polarization of the initial electron and
positron beams, it is more easy to detect the possible signals of
the new gauge bosons $B_{H}$ and $Z_{H}$ in the future $LC$
experiments.

\vspace{0.5cm}
\begin{figure}[htb]
\vspace{-1cm}
\begin{center}
\epsfig{file=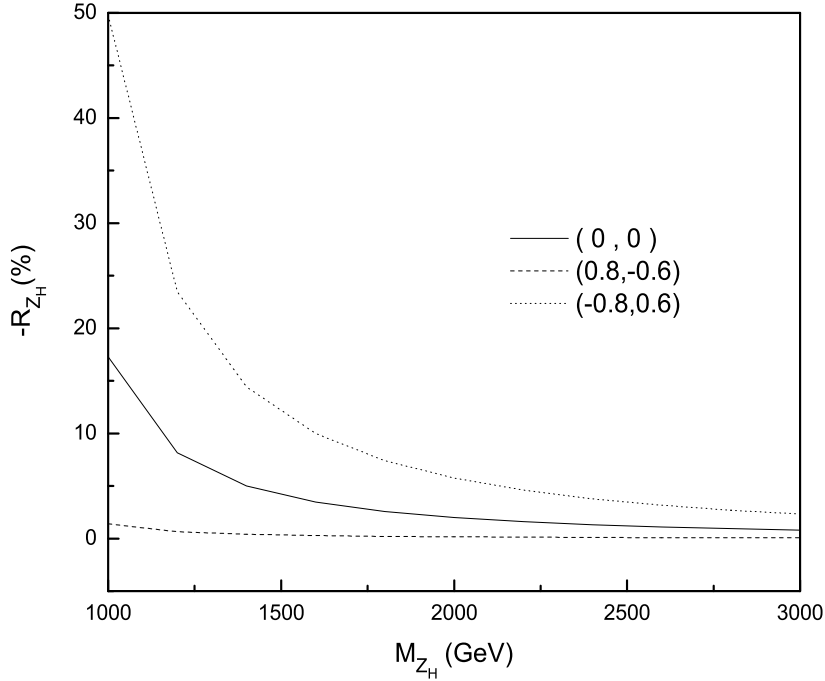,width=350pt,height=300pt} \vspace{-1cm}
\hspace{1cm} \caption{The relative correction parameter
$R_{Z_{H}}$ as a function of the $Z_{H}$ mass $M_{Z_{H}}$ for
\hspace*{1.8cm} $c=0.3$ and $(P_{e},P_{\bar{e}})=(0,0),
(0.8,-0.6), (-0.8,0.6)$.} \label{ee}
\end{center}
\end{figure}

\noindent{\bf IV. Gauge bosons $B_{H}$, $Z_{H}$ and the
forward-backward asymmetry $A_{FB}(t\bar{t})$}

The events generated by the process $e^{+}e^{-}\rightarrow
f\bar{f}$ can be characterized by the  momentum direction of the
emitted fermion. If we assume that the final state fermion travels
forward(F) or backward(B) with respect to the electron beam, than
the forward-backward asymmetry can be defined as:
\begin{equation}
A_{FB}=\frac{\sigma_{F}-\sigma_{B}}{\sigma_{F}+\sigma_{B}},
\end{equation}
which is easier to be measured because only the identification of
the charge of the fermion and the measurement of its direction are
needed [13]. It can be measured for all tagged flavors and
inclusively for hadrons. Thus, it is needed to calculate the
contributions of $B_{H}$ exchange and $Z_{H}$ exchange to the
forward-backward asymmetry $A_{FB}(t\bar{t})$.

\begin{figure}[htb]
\vspace{0cm}
  \centering
   \includegraphics[width=3.3in]{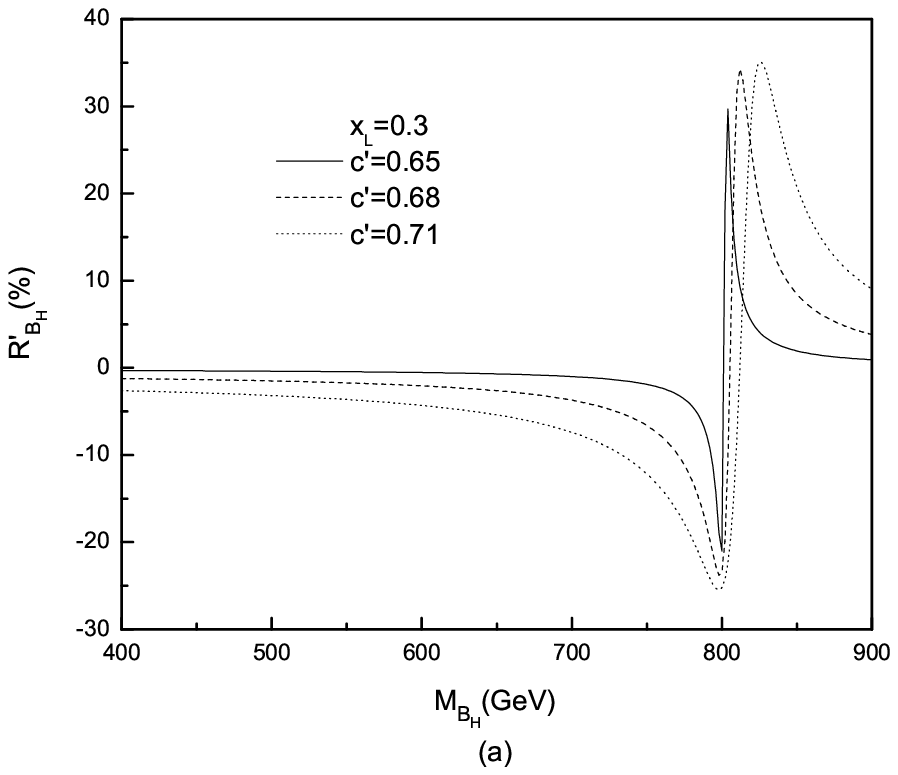}
    \hspace{-0.4in}
   \includegraphics[width=3.3in]{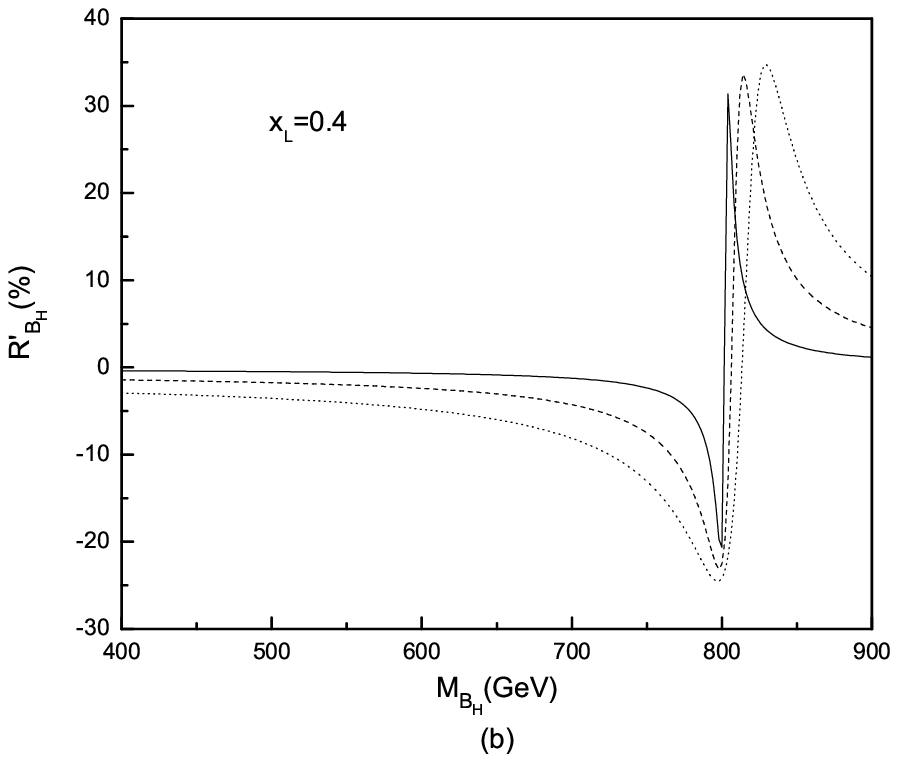}
\end{figure}

\begin{figure}[htb]
\vspace{-1.5cm}
  \centering
   \includegraphics[width=3.3in]{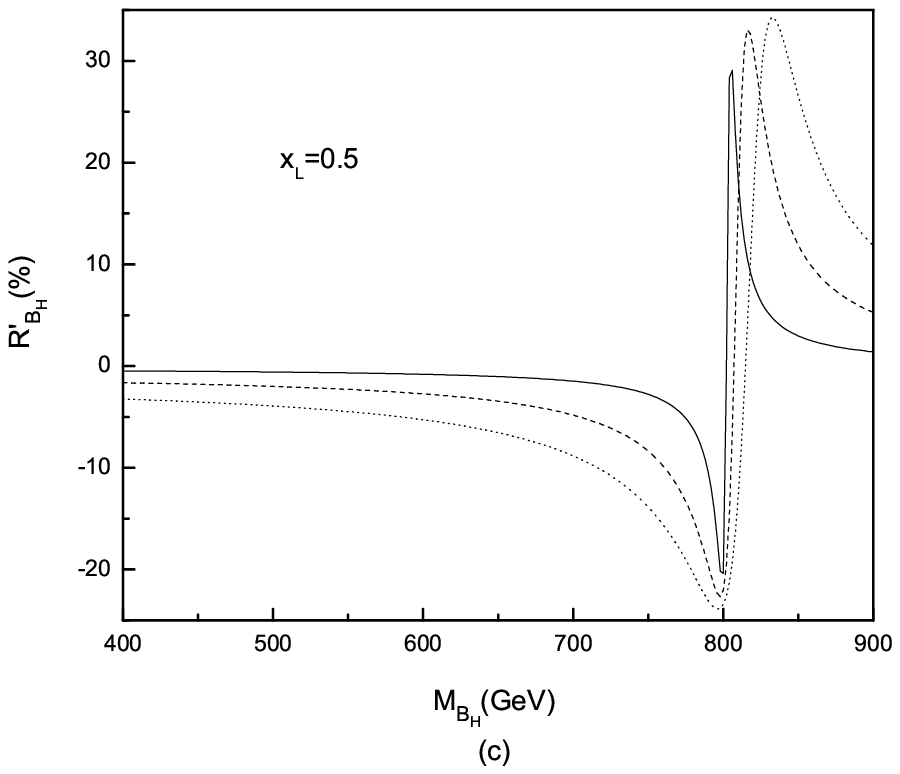}
    \hspace{-0.4in}
   \includegraphics[width=3.3in]{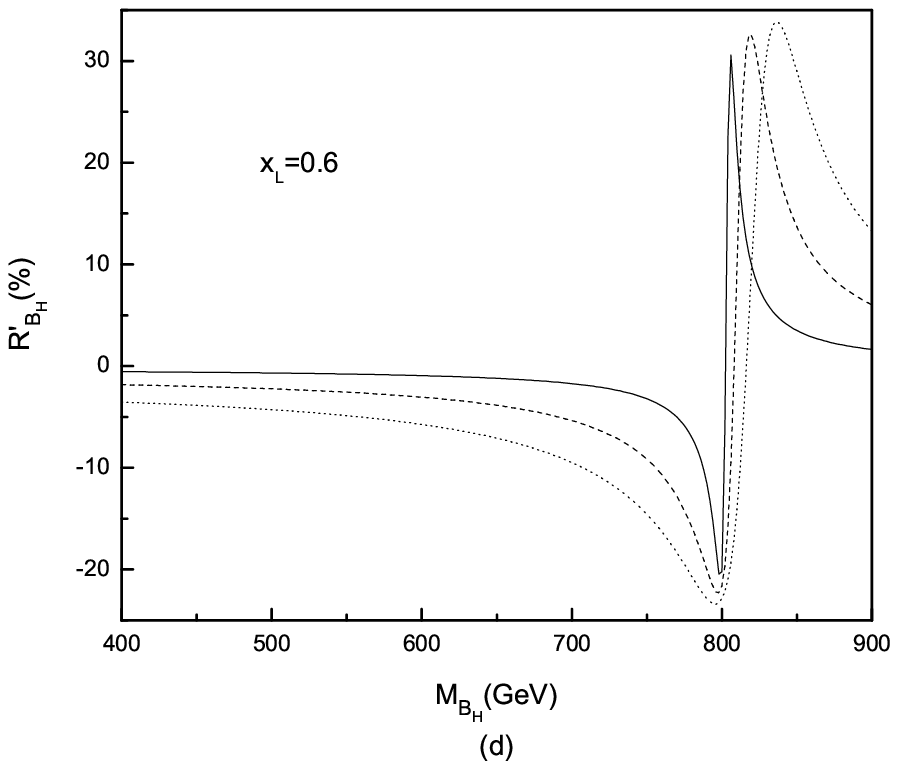}
  \caption{The relative correction parameter
$R'_{B_{H}}$ as a function of  $M_{B_{H}}$ for different values
\hspace*{1.8cm} of the mixing parameters $c'$ and $x_{L}$.}
\label{ee}
\end{figure}

The total formula of $A_{FB}(t\bar{t})$ for the new gauge bosons
$B_{H}$ and $Z_{H}$ including the contributions of the $SM$ gauge
bosons $\gamma$ and $Z$ can be written as:
\begin{equation}
A_{FB}^{B_{H}}(t\bar{t})=\frac{M_{2}^{B_{H}}(t\bar{t})}{M_{1}^{B_{H}}(t\bar{t})},
\hspace{1.5cm}
A_{FB}^{Z_{H}}(t\bar{t})=\frac{M_{2}^{Z_{H}}(t\bar{t})}{M_{1}^{Z_{H}}(t\bar{t})},
\end{equation}
where
\begin{eqnarray}
M_{2}^{B_{H}}(t\bar{t})&=&\beta\{\frac{2e^{2}}{3}g_{A}^{Zee}g_{A}^{Ztt}
\frac{4S(M_{Z}^{2}-S)}{(S-M_{Z}^{2})^{2}+M_{Z}^{2}\Gamma_{Z}^{2}}\nonumber\\
&&+\frac{2e^{2}}{3}g_{A}^{B_{H}ee}g_{A}^{B_{H}tt}\frac{4S(M_{B_{H}}^{2}-S)}{(S-M_{B_{H}}
^{2})^{2}
+M_{B_{H}}^{2}\Gamma_{B_{H}}^{2}}\nonumber\\&&+g_{V}^{Zee}g_{A}^{Zee}g_{V}^{Ztt}g_{A}^{Ztt}
\frac{8S^{2}}{(S-M_{Z}^{2})^{2}+M_{Z}^{2}\Gamma_{Z}^{2}}\nonumber\\
&&+g_{V}^{B_{H}ee}g_{A}^{B_{H}ee}g_{V}^{B_{H}tt}g_{A}^{B_{H}tt}
\frac{8S^{2}}{(S-M_{B_{H}}^{2})^{2}+M_{B_{H}}^{2}\Gamma_{B_{H}}^{2}}\nonumber\\
&&+(g_{V}^{Zee}g_{A}^{B_{H}ee}+g_{V}^{B_{H}ee}g_{A}^{Zee})(g_{V}^{Ztt}g_{A}^{B_{H}tt}
+g_{A}^{Ztt}g_{V}^{B_{H}tt})\nonumber\\&&\frac{4S^{2}[(S-M_{Z}^{2})(S-M_{B_{H}}^{2})+M_{Z}
\Gamma_{Z}M_{B_{H}}\Gamma_{B_{H}}]}{[(S-M_{Z}^{2})^{2}+M_{Z}^{2}\Gamma_{Z}^{2}]
[(S-M_{B_{H}}^{2})^{2}+M_{B_{H}}^{2}\Gamma_{B_{H}}^{2}]}\},\\
M_{1}^{B_{H}}(t\bar{t})&=&\{\frac{16e^{4}}{9}(1-\frac{\beta^{2}}{3})
+\frac{8e^{2}}{3}(1-\frac{\beta^{2}}{3})g_{V}^{Zee}g_{V}^{Ztt}
\frac{2S(M_{Z}^{2}-S)}{(S-M_{Z}^{2})^{2}+M_{Z}^{2}\Gamma_{Z}^{2}}\nonumber\\
&&[(g_{V}^{Zee})^{2}+(g_{A}^{Zee})^{2}]
[4(1-\frac{\beta^{2}}{3})[(g_{V}^{Ztt})^{2}+(g_{A}^{Ztt})^{2}]
-4(1-\beta^{2})(g_{A}^{Ztt})^{2}]\times\nonumber\\&&\frac{S^{2}}{(S-M_{Z}^{2})^{2}+M_{Z}^{2}
\Gamma_{Z}^{2}}+\frac{8e^{2}}{3}(1-\frac{\beta^{2}}{3})g_{V}^{B_{H}ee}
g_{V}^{B_{H}tt}\frac{2S(M_{B_{H}}^{2}-S)}{(S-M_{B_{H}}^{2})^{2}+M_{B_{H}}^{2}
\Gamma_{B_{H}}^{2}}\nonumber\\&&+[(g_{V}^{B_{H}ee})^{2}+(g_{A}^{B_{H}ee})^{2}]
[4(1-\frac{\beta^{2}}{3})[(g_{V}^{B_{H}tt})^{2}+(g_{A}^{B_{H}tt})^{2}]-4(1-\beta^{2})
(g_{A}^{B_{H}tt})^{2}]\nonumber\\&&\frac{S^{2}}
{(S-M_{B_{H}}^{2})^{2}+M_{B_{H}}^{2}\Gamma_{B_{H}}^{2}}+(g_{V}^{Zee}
g_{V}^{B_{H}ee}+g_{A}^{Zee}g_{A}^{B_{H}ee})\nonumber\\&&[4(1-\frac{\beta^{2}}{3})
(g_{V}^{Ztt}g_{V}^{B_{H}tt}+g_{A}^{Ztt}g_{A}^{B_{H}tt})-4(1-\beta^{2})(g_{A}^{B_{H}tt})
(g_{A}^{Ztt})]\nonumber\\&&\frac{2S^{2}[(S-M_{Z}^{2})(S-M_{B_{H}}^{2})+M_{Z}
\Gamma_{Z}M_{B_{H}}\Gamma_{B_{H}}]}{[(S-M_{Z}^{2})^{2}+M_{Z}^{2}\Gamma_{Z}^{2}]
[(S-M_{B_{H}}^{2})^{2}+M_{B_{H}}^{2}\Gamma_{B_{H}}^{2}]}\},\\
M_{2}^{Z_{H}}(t\bar{t})&=&\beta\{\frac{2e^{2}}{3}g_{A}^{Zee}g_{A}^{Ztt}
\frac{4S(M_{Z}^{2}-S)}{(S-M_{Z}^{2})^{2}+M_{Z}^{2}\Gamma_{Z}^{2}}\nonumber\\
&&+\frac{2e^{2}}{3}g_{A}^{Z_{H}ee}g_{A}^{Z_{H}tt}\frac{4S(M_{Z_{H}}^{2}-S)}{(S-
M_{Z_{H}}^{2})^{2}
+M_{Z_{H}}^{2}\Gamma_{Z_{H}}^{2}}\nonumber\\&&+g_{V}^{Zee}g_{A}^{Zee}g_{V}^{Ztt}g_{A}^{Ztt}
\frac{8S^{2}}{(S-M_{Z}^{2})^{2}+M_{Z}^{2}\Gamma_{Z}^{2}}\nonumber\\
&&+g_{V}^{Z_{H}ee}g_{A}^{Z_{H}ee}g_{V}^{Z_{H}tt}g_{A}^{Z_{H}tt}
\frac{8S^{2}}{(S-M_{Z_{H}}^{2})^{2}+M_{Z_{H}}^{2}\Gamma_{Z_{H}}^{2}}\nonumber\\
&&+(g_{V}^{Zee}g_{A}^{Z_{H}ee}+g_{V}^{Z_{H}ee}g_{A}^{Zee})(g_{V}^{Ztt}g_{A}^{Z_{H}tt}
+g_{A}^{Ztt}g_{V}^{Z_{H}tt})\nonumber\\&&\frac{4S^{2}[(S-M_{Z}^{2})(S-M_{Z_{H}}^{2})+M_{Z}
\Gamma_{Z}M_{Z_{H}}\Gamma_{Z_{H}}]}{[(S-M_{Z}^{2})^{2}+M_{Z}^{2}\Gamma_{Z}^{2}]
[(S-M_{Z_{H}}^{2})^{2}+M_{Z_{H}}^{2}\Gamma_{Z_{H}}^{2}]}\},\\
M_{1}^{Z_{H}}(t\bar{t})&=&\{\frac{16e^{4}}{9}(1-\frac{\beta^{2}}{3})
+\frac{8e^{2}}{3}(1-\frac{\beta^{2}}{3})g_{V}^{Zee}g_{V}^{Ztt}
\frac{2S(M_{Z}^{2}-S)}{(S-M_{Z}^{2})^{2}+M_{Z}^{2}\Gamma_{Z}^{2}}\nonumber\\
&&[(g_{V}^{Zee})^{2}+(g_{A}^{Zee})^{2}]
[4(1-\frac{\beta^{2}}{3})[(g_{V}^{Ztt})^{2}+(g_{A}^{Ztt})^{2}]
-4(1-\beta^{2})(g_{A}^{Ztt})^{2}]\times\nonumber\\&&\frac{S^{2}}{(S-M_{Z}^{2})^{2}+M_{Z}^{2}
\Gamma_{Z}^{2}}+\frac{8e^{2}}{3}(1-\frac{\beta^{2}}{3})g_{V}^{Z_{H}ee}
g_{V}^{Z_{H}tt}\frac{2S(M_{Z_{H}}^{2}-S)}{(S-M_{Z_{H}}^{2})^{2}+M_{Z_{H}}^{2}
\Gamma_{Z_{H}}^{2}}\nonumber\\&&+[(g_{V}^{Z_{H}ee})^{2}+(g_{A}^{Z_{H}ee})^{2}]
[4(1-\frac{\beta^{2}}{3})[(g_{V}^{Z_{H}tt})^{2}+(g_{A}^{Z_{H}tt})^{2}]-4(1-\beta^{2})
(g_{A}^{Z_{H}tt})^{2}]\nonumber\\&&\frac{S^{2}}
{(S-M_{Z_{H}}^{2})^{2}+M_{Z_{H}}^{2}\Gamma_{Z_{H}}^{2}}+(g_{V}^{Zee}
g_{V}^{Z_{H}ee}+g_{A}^{Zee}g_{A}^{Z_{H}ee})\nonumber\\&&[4(1-\frac{\beta^{2}}{3})
(g_{V}^{Ztt}g_{V}^{Z_{H}tt}+g_{A}^{Ztt}g_{A}^{Z_{H}tt})-4(1-\beta^{2})(g_{A}^{Z_{H}tt})
(g_{A}^{Ztt})]\nonumber\\&&\frac{2S^{2}[(S-M_{Z}^{2})(S-M_{Z_{H}}^{2})+M_{Z}
\Gamma_{Z}M_{Z_{H}}\Gamma_{Z_{H}}]}{[(S-M_{Z}^{2})^{2}+M_{Z}^{2}\Gamma_{Z}^{2}]
[(S-M_{Z_{H}}^{2})^{2}+M_{Z_{H}}^{2}\Gamma_{Z_{H}}^{2}]}\}.
\end{eqnarray}
In above equations, we have assumed that the initial electron and
positron beams are not polarized.

To see whether the new gauge bosons $B_{H}$ and $Z_{H}$ can
produce significant deviations from the $SM$ prediction value for
$A_{FB}(t\bar{t})$, we plot the relative correction parameters
$R'_{B_{H}}=\delta A_{FB}^{B_{H}}(t\bar{t})/A_{FB}^{SM}(t\bar{t})$
and $R'_{Z_{H}}=\delta
A_{FB}^{Z_{H}}(t\bar{t})/A_{FB}^{SM}(t\bar{t})$ as functions of
$M_{B_{H}}$ and $M_{Z_{H}}$ in Fig.5 and Fig.6, respectively. From
Fig.5 and Fig.6 we can see that, in most of the parameter space
preferred by the electroweak precision data, the absolute values
of the relative correction parameters $R'_{B_{H}}$ and
$R'_{Z_{H}}$ are smaller than $5\%$. The absolute values of
$R'_{Z_{H}}$ is larger than $5\%$ only for the mixing parameter
$c=0.5$ and $1TeV\leq M_{Z_{H}}\leq 1.4TeV$. $B_{H}$ exchange
makes the deviation of the forward-backward asymmetry
$A_{F_{B}}(t\bar{t})$ from its $SM$ value may be positive or
negative, which depends on the $B_{H}$ mass $M_{B_{H}}$. The
resonance peak can emerge for $M_{B_{H}}\approx800GeV$.
Furthermore, the absolute value of $R'_{B_{H}}$ increases as the
mixing parameters $c'$ and $x_{L}$ increasing. For $c'\geq0.71$,
$x_{L}\geq 0.5$, and $600GeV\leq M_{B_{H}}\leq 1000GeV$, the
absolute value of $R'_{B_{H}}$ is larger than $5\%$, which might
be detected in the future $LC$ experiments. However, for
$c'\leq0.68$ and $x_{L}\leq 0.4$, except for a small region near
$M_{B_{H}}=800GeV$, the absolute value of $R'_{B_{H}}$ is smaller
than $5\%$.

Similar to above calculation, we can obtain the corrections of
$B_{H}$ exchange and $Z_{H}$ exchange to the forward-backward
symmetry $A_{FB}(f\bar{f})$ with $f$=$\mu,$ $\tau,$ $b,$ or, $c$.
From the coupling formula of the new gauge bosons $B_{H}$ and
$Z_{H}$ to differently fermions given in Ref.[7], we can surmise
that the conclusions are similar to those for $A_{FB}(t\bar{t})$.
We have confirmed this expectation through explicit calculation.
Certainly, the contributions of $B_{H}$ exchange to
$A_{FB}(f\bar{f})$ mainly dependent on the free parameters
$M_{B_{H}}$ and $c'$, while the contributions of $B_{H}$ exchange
to $A_{FB}(t\bar{t})$ mainly dependent on the free parameters
$M_{B_{H}}$, $c'$, and $x_{L}$.

\vspace{0.5cm}
\begin{figure}[htb]
\vspace{-1cm}
\begin{center}
\epsfig{file=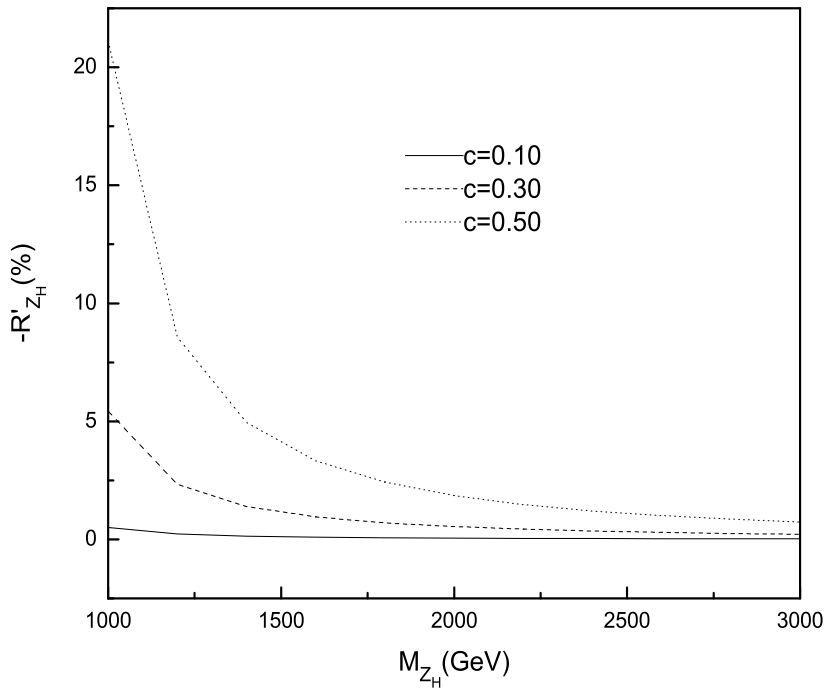,width=350pt,height=300pt} \vspace{-1cm}
\hspace{1cm} \caption{The relative correction parameter
$R'_{Z_{H}}$ as a function of $M_{Z_{H}}$ for three values of
\hspace*{1.8cm} the mixing parameter $c$.} \label{ee}
\end{center}
\end{figure}

\noindent{\bf V. Conclusions and discussins}

An $LC$ will be an ideal machine for precisely testing the $SM$
and probing $NP$ beyond the $SM$. Some kinds of $NP$ predict the
existence of new particles that will be manifested as a rather
spectacular resonance in the $LC$ experiments if the achievable
c.m. energy $\sqrt{S}$ is sufficient. Even if their masses exceed
the c.m. energy $\sqrt{S}$, the $LC$ experiments also retain an
indirect sensitivity through a precision study their virtual
corrections to observables.

It is widely believed that the top quark, with a mass of the order
of the electroweak scale, will be a sensitive probe into $NP$
beyond the $SM$. The quantum correction effects of the new
particles to some $SM$ processes involving top quark are more
important than those for lighter fermions. Thus, the top quark
plays a key role in the quest for deviations of observables from
their $SM$ predictions. On the other hand, top quark pairs can be
copiously produced mainly through the process
$e^{+}e^{-}\rightarrow t\bar{t}$ in the future $LC$ experiments.
So, in this paper, we discuss and calculate the corrections of the
new gauge bosons $B_{H}$ and $Z_{H}$ predicted by the $LH$ model
to the production cross section $\sigma(t\bar{t})$ and the
forward-backward asymmetry $A_{FB}(t\bar{t})$ of the process
$e^{+}e^{-}\rightarrow t\bar{t}$.

The $LH$ model has all essential features of the little Higgs
models. So, in this paper, we give our numerical results in the
context of the $LH$ model, although many alternatives have been
proposed [2,3]. We find that the new gauge bosons $Z_{H}$ and
$B_{H}$ can produce significant correction effects on the process
$e^{+}e^{-}\rightarrow t\bar{t}$, which can be further enhanced by
the suitably polarized beams. In most of the parameter space
$f=1TeV\sim 2TeV, c'=0.62\sim0.73, c=0.1\sim0.5,$ and
$x_{L}=0.3\sim0.6$, which consistent with the electroweak
precision data, the absolute value of the relative correction
parameter $R_{B_{H}}$ generated by $B_{H}$ exchange is larger than
5\%. As long as $1TeV\leq M_{Z_{H}}\leq 1.5TeV,$ and $0.3\leq
c\leq 0.5,$ the absolute value of $R_{Z_{H}}$ is larger than 5\%.
Thus, we can say that, with reasonable values of the parameters in
the $LH$ model, the possible signals of the new gauge bosons
$B_{H}$ and $Z_{H}$ can be detected via the process
$e^{+}e^{-}\rightarrow t\bar{t}$ in the future $LC$ experiments
with the c.m. energy $\sqrt{S}=800GeV$. However, $B_{H}$ exchange
and $Z_{H}$ exchange can only generate very small corrections to
the forward-backward asymmetry $A_{FB}(t\bar{t})$ in most of the
parameter space. It is possible that, in very small range of the
parameter space, the possible signals of $B_{H}$ and $Z_{H}$ might
be detected via measuring the deviations of $A_{FB}(t\bar{t})$
from its $SM$ prediction.

The couplings of the new gauge boson $B_{H}$ to fermions are quite
model dependent, which depend on the choice of the fermion $ U(1)
$ charges under the two $ U(1)$ groups. The $ U(1) $ charges of
the SM fermions are constrained  by requiring that the Yukawa
couplings are gauge invariant and maintaining the usual SM
hypercharge assignment. Combing the gauge invariance of the Yukawa
couplings with the $ U(1)$ anomaly-free can fix all of the $U(1)$
charge values. In this paper, we have used the couplings of the
$B_{H}$ to fermions, which come from this kind of choice.
Certainly, this is only one example of all possible $U(1)$ charge
assignments. In other little Higgs models, several alternatives
for the $U(1)$ charge choice exist[2, 3, 10], the numerical
results for the new gauge boson $B_{H}$ obtained in this paper
might be changed.

\vspace{1.0cm} \noindent{\bf Acknowledgments}

This work was supported in part by the National Natural Science
Foundation of China under the grant No.90203005 and No.10475037
and the Natural Science Foundation of the Liaoning Scientific
Committee(20032101).

\null

\end{document}